\def\ba{\bar a}

\def\be{\begin{equation}}
\def\ee{\end{equation}}
\def\te{\end{equation}}
\def\bea{\begin{eqnarray}}
\def\ba{\begin{eqnarray}}
\def\eea{\end{eqnarray}}
\def\ea{\end{eqnarray}}
\def\tea{\end{eqnarray}}
\def\nn{\nonumber}

\def\lbr{\left \langle}
\def\rbr{\right\rangle }

\def\rmd{\text{d}}

\def\rme{\text{e}}
\def\tr{\text{tr}}
\def\gr{g_{ret}}

\def\g{\raisebox{.4ex}{$\gamma$}}

\def\s{\sigma}

\def\cP{{\cal P}}
\def\cW{{\cal W}}

\def\(#1){(\ref{#1})}

\newskip\humongous \humongous=0pt plus 1000pt minus 1000pt

\newif\ifdtup

\documentclass[pre,aps,nofootinbib]{revtex4}

\usepackage{amsmath}

\begin{document}

\bibliographystyle{apsrev}

\title{Quantum and classical fluctuation theorems\\ from a decoherent
histories,
open-system analysis}

\author{Y. Suba\c{s}\i,  and  B. L. Hu}
\affiliation{Maryland Center for Fundamental Physics and Joint Quantum
Institute,\\
University of Maryland, College Park, Maryland 20742, USA}

\date{\today}

\begin{abstract}
In this paper we present a first-principles analysis of the
nonequilibrium work distribution and the free energy difference of a
quantum system interacting with a general environment (with arbitrary
spectral density and for all temperatures) based on a well-understood
micro-physics (quantum Brownian motion) model under the conditions
stipulated by the Jarzynski equality [C. Jarzynski, Phys. Rev. Lett.
78, 2690 (1997)] and Crooks' fluctuation theorem [G. E. Crooks,
Phys. Rev. E 60, 2721 (1999)] (in short FTs). We use the decoherent history
conceptual framework to explain how the notion of trajectories in a
quantum system can be made viable and use the environment-induced
decoherence scheme to assess the strength of noise which could provide
sufficient decoherence to warrant the use of trajectories to define
work in open quantum systems. From the solutions to the Langevin
equation governing the stochastic dynamics of such systems we were
able to produce formal expressions for these quantities entering in
the FTs,  and from them prove explicitly the validity of the FTs at
the high temperature limit.  At low temperatures our general results
would enable one to identify the range of parameters where FTs may not hold or
need be expressed differently. We
explain the relation between classical and quantum FTs and the advantage of
this micro-physics open-system approach over the  phenomenological modeling and
energy-level calculations for substitute closed quantum systems.
\end{abstract}

\maketitle

\section{Introduction}

Unlike in equilibrium statistical physics, few theorems of generality
are established for nonequilibrium systems. Hence any valid statement
with a broad spectrum of implications and wide range of applications
is of great value. The fluctuation theorems (FTs) of Jarzynski
\cite{Jarzynski1997a} and Crooks \cite{Crooks1999} \footnote{By the FTs we
refer specifically to Jarzynski's equality and
Crooks's fluctuation theorem only.} in nonequilibrium
statistical mechanics are of such a nature which have
stimulated intense research interest and activities in the past
decade. 
For earlier work on entropy fluctuation theorems, such as by Cohen, Evans,
Searles and others, see, e.g.,\cite{EvansSearles2002}.
FTs relate some  equilibrium thermodynamic quantities of a
physical system, like free energy differences, to the averages of
mechanical quantities in nonequilibrium processes, like exponentiated
work. For complex biological systems like proteins and DNAs the free
energy differences are difficult to calculate while the averages of
work in nonequilibrium processes can be obtained from measurements in
experiments or via numerical simulations.

\subsection{Background and Basic Issues}

\paragraph{How to define work in quantum physics?}
\label{sec:howtodefinework}
The fluctuation theorems were originally derived for classical
thermodynamic systems. It is natural to ask  if they hold for quantum systems,
and if not, under what
conditions would they fail, and whether there exist quantum fluctuation
theorems (QFTs) different in form and content from the classical FTs.
If this is not possible, can one find an approximate form in terms of
corrections to the classical FTs.
To our knowledge corrections have been derived  but a full QFT that
is valid for all conditions (e.g., non-Ohmic spectral density of the
environments at low temperature) is still at large.

In these endeavors the main conceptual obstacle is how to make sense
of work in a quantum setting.  To begin with, work is not an observable
\cite{Talkner2007}, and as such,
treating it as a quantum mechanical operator \cite{Chernyak2004, Monnai2005} is
largely a computational
convenience. Thus the foremost task is to find a physically meaningful
definition and an operationally feasible way to calculate it. We will address
this issue with a new approach described below.

Let us try to appreciate the content of this pivotal point.
In classical mechanics \textit{exclusive work} \cite{Jarzynski2007} imparted to
a system, say a particle, is
defined as the integration of applied force on the system with
displacement along a path. The force is exerted by an external agent
which causes the system to move along a trajectory. Once one knows the
trajectory, work can be calculated, but the difficulty for quantum
system is that particles don't follow trajectories, they are
described by a wave function which is a very different notion and
entity from paths. The key challenge is to make sense of trajectories in
quantum physics. We mention several approaches below and then present our own.

\paragraph{Closed versus  open quantum system}

If one restricts one’s attention to closed quantum systems, i.e., isolated
quantum systems having no interaction with any of their environments, one can
define work via transitions between the system’s energy levels (quantum jumps)
\cite{Tasaki2000, Campisi2011a}, and general agreement seems to be reached.
However this is merely an idealization of realistic physical
systems which are more often open. The influence of their environments which
the system of
interest interacts with need be accounted for in the open system's evolution.
Even in the simplest
cases when one talks about temperature or refers to (equilibrium)
thermodynamic quantities a heat bath (canonical ensemble) or a
particle reservoir (grand canonical ensemble) is implicitly assumed,
which are open-system setups.

Since for closed quantum systems fluctuation theorems can be easily
derived, one can think of the system + environment as a closed system and work
out the QFTs. This was done in
\cite{Talkner2009, Campisi2011a, Monnai2010}. However this formulation has the
innate
shortcoming that the work defined therein  requires the change in energy of the
\emph{combined} system while the FTs refer to the work on the \emph{open}
system of interest (being a subsystem of the combined, whose dynamics includes
the back-action from the other subsystems as its environment).

\paragraph{Microscopic models for open system dynamics}

The use of a microphysics model such as the quantum Brownian motion (QBM) model
described below could provide a rigorous basis for any phenomenological
description. It makes explicit any assumption made in the phenomenological
models which enables one to clearly define the range of validity of the results
derived from each model, as well as being able to provide the details in the
derivations with or without the corresponding assumptions. Applying methods of
nonequilibrium statistical
mechanics such as the Zwangzig-Mori-Nakajima projection operator or the
Feynman-Vernon influence functional (IF) formalism to a microscopic model
consummates the objectives of  quantum open-system treatment. Using these
methods one obtains a description of the open-system dynamics in terms of open
system
variables alone, while the dynamics of the open system already factored in  the
back-action of the environment. Environment related quantities like heat can
also be addressed within this framework.  The IF is the method we have used
before and prefer, since it has the advantage of including the back-action in a
self-consistent manner and one can invoke field theory techniques (by way of
the almost equivalent Schwinger-Keldysh closed time path formalism) to address
nonequilibrium statistical mechanics issues.

Using microscopic models and open system dynamics several suggestions for
trajectories have been made. For example, De Roeck \cite{Deroeck2007} used the
unraveling of the open system master equation and compared his results to that
of the closed system approach.  Deffner et al. \cite{Deffner2011a} used the
quantum Smoluchowski equation (QSE), which was derived from taking the  high
friction and high temperature limit in \cite{Dillenschneider2009},
as a starting point.  They considered the solution to the QSE in terms of
classical path integrals and
interpreted these paths as trajectories. But these trajectories are difficult
to interpret physically, being more in the nature as devices (to help solve a
differential equation) than actual physical entities.
By making the assumption that the reduced dynamics of a driven open quantum
subsystem is described by a quantum master equation Esposito and Mukamel
\cite{Esposito2006} recast its solution in a representation which takes the
form of a birth-death master equation (BDME) with time-dependent rates and used
it to define ``quantum" trajectories.  But these QSE and BDME, just as the
Pauli master equation, govern transition probabilities, are equivalent to a 
reduced density matrix with only diagonal elements, and thus contain no quantum
phase information \footnote{This may be viewed as the completely decohered end
product of a decoherent history or environment-induced decoherence process
(complete diagonalization of the reduced density matrix)
but as we shall explain in more detail below, it corresponds to the case of
very strong noise acting on the subsystem, which is possible for high
temperatures, and thus it falls under the parameter regime where the classical
FTs are valid. In fact for Gaussian systems, the QFT derived under these
conditions have exactly the same form as the classical FTs.}.

Alternatively Crooks \cite{Crooks2008a} proved the Jarzynski equality by
considering the Markovian dynamics of a quantum system in the following
setting: Instead of measuring the system, generalized measurement
superoperators were used to represent measurements of heat flow. If the quantum
environment is assumed to be large, to have rapidly decohered and always remain
at thermal equilibrium, plus being uncorrelated and unentangled with the
system, then the change in energy of the bath can be measured without further
disturbing the dynamics of the system. 

In comparison with earlier work our approach is closest in spirit to that of
Chernyak and Mukamel \cite{Chernyak2004}. However our methods (they use
superoperators in Liouville space) and interpretations (they use von Neumann's
wave function collapse for quantum measurement) are different.  We will detail
the differences after we have a chance to describe our approach.

\subsection{Our approach and findings}

For the sake of conciseness we just state what we do and name the ingredients
in our approach here,  leaving more detailed explanations to the next section.

In  this paper we analyze  the fluctuation theorems (FTs) using the
exactly solvable microscopic quantum Brownian motion (QBM) model of a quantum
harmonic oscillator coupled to a heat
bath of N quantum harmonic oscillators with arbitrary spectral density
function and for all temperatures. This is referred to as a `general'
environment in \cite{Hu1992} where an exact master equation for these
full ranges was obtained and where our discussions in the application
of this model to QFTs are based upon \footnote{We advise against
calling this a \textit{non-Markovian environment}, because non-Markovian refers
to stochastic \emph{processes}, not systems. Instead, use, e.g., 
\textit{colored noise environments}, which can engender non-Markovian dynamics
in the open subsystem.}. The low temperature results are of special interest
for the derivation of the QFT since it measures the deviation from the
classical FTs.

\paragraph{Decoherent history approach to define trajectories for quantum
systems}

We resort to the conceptual framework of decoherent \cite{Gell-Mann1993}
or consistent \cite{Griffiths1984, Omnes1992} histories (dechis) 
and the key notion of decoherence for understanding the process of quantum to
classical transition.  We believe
this is the most faithful and intuitive way of defining trajectories
or explaining how they arise from quantum mechanics. To be more
precise, these trajectories are actually stochastic classical paths in
a quasi-classical domain as a result of decoherence in the histories.
They arise by the action of noise which are defined as variations in
neighboring histories. (For a succinct explanation of the first point
see e.g., \cite{Griffiths1984,Omnes1988} and \cite{Gell-Mann1990} on the second
point.)

\paragraph{Environment-induced decoherence for explicit computations}

While the decoherent history paradigm is conceptually clear for
explaining the origin and mechanisms in the emergence of classical
stochastic  trajectories, it is less versatile in actual computations.
The environment-induced decoherence (\emph{envdec}) approach can be of more
practical use. Here, the approximate diagonalization of the reduced
density matrix of the reduced or open system with respect to some
basis is used as a signifier of decoherence of the quantum system in
transit to classicality, whereby the notion of trajectory becomes
viable. But which basis? This is the physically relevant issue. The
quantum system is more readily decohered in the so-called ``pointer
basis"  \cite{Zurek1982} which is affected by the form of interaction
between the system and its environment. Here, with an
explicit environment specified, it is easier to see how noise arises
and its nature (colored, multiplicative \cite{Hu1993}) than in the
decoherent history approach.  The connection between these two
approaches is discussed in \cite{PazZurek1993}. An explicit model calculation
(the QBM model) was given in the {\it dechis} approach \cite{Dowker1992} where
one can compare these two approaches in operational details.

\paragraph{Significance of stochastic regime between quantum and classical}

In reference to trajectories of quantum origin we notedly attach the
word `stochastic' to classical. This is because there is a stochastic
component to them after the quantum histories decohere. They are
described by a probability distribution function. Each such trajectory
is a realization of this distribution. 
Taking the stochastic average of an ensemble of such trajectories will yield
the unique classical path which is a solution of a deterministic classical
equation of motion.

Decoherence is due to noise, quantum or thermal or both. In the {\it envdec}
scheme, one can see this explicitly from the stochastic equations
governing the open (reduced) system. Noise is responsible for quantum
diffusion which brings forth decoherence. The stronger the noise the
more complete the decoherence process and the more classical the
trajectories.  In fact for the QBM model there are two diffusion
terms: a normal diffusion dominates at high temperatures and an
anomalous diffusion which dominates at low temperature. The latter is
what one should focus on in marking the difference between the
classical and the quantum FTs. Therefore the behavior of a system in
the stochastic regime actually holds the key to quantum-classical
transition or correspondence. It is particularly suitable for the
exploration of FTs in open systems as they are also cast in a stochastic
framework in terms of the probability distribution of work.

\paragraph{Our findings} In this paper we present a first-principles
analysis of the nonequilibrium work distribution and the free energy
difference of a quantum system interacting with a general environment
(with arbitrary spectral density and for all temperatures) based on a
well-understood micro-physics (quantum Brownian motion) model under
the conditions stipulated by the Jarzynski equality  and Crooks'
fluctuation theorem (FTs). We use the decoherent history conceptual
framework to explain how the notion of trajectories in a quantum
system can be made viable and use the environment-induced decoherence
scheme to assess the strength of noise which could provide sufficient
decoherence to warrant the use of trajectories to define work in open
quantum systems. From the solutions to the Langevin equation governing
the stochastic dynamics of such systems we were able to produce formal
expressions for these quantities entering in the FTs,  and from them
prove explicitly the validity of the FTs at the high temperature
limit.  At low temperatures our general results would enable one to
identify the range of parameters where FTs may not hold or
need be expressed differently. We explain the
relation between classical and quantum FTs and the advantage of this
micro-physics open-system approach over the  phenomenological modeling and
energy-level calculations for substitute closed quantum systems.

\section {Key points and main ideas}

Because we are seeking A) a derivation of quantum fluctuation
theorems in nonequilibrium physics by applying concepts and
practices in B) quantum foundation and measurement theory via decoherent
histories and 
environment-induced decoherence with its ensuing classical stochastic equations
it might be useful to give a
brief summary of the key ideas and procedures in this section for ease
of cross-reference. For good reviews on these subjects we mention
\cite{Campisi2011a} for A) and \cite{Omnes1992,Griffiths1984,Zurek2003} for B).
Readers familiar with A) can skip the first subsection, readers familiar with
B) can skip the last subsection. Readers familiar with both please go to the
next section.

\subsection{Fluctuation Theorems}

\subsubsection{Classical FTs}

We briefly review the premises of these theorems in classical
Hamiltonian dynamics \footnote{For a recent review on classical FTs see
\cite{Jarzynski2011}.}. 
Consider a classical system, whose dynamics is governed by the Hamiltonian
$H(\lambda_t)$. $\lambda$ is a deterministic parameter with a prescribed time
dependence. At some initial time $t_0$ which without loss of generality can  be
taken to be zero, the
system is prepared in a thermal state $ \exp(-\beta
H(\lambda_0))/Z_0$. Then the parameter $\lambda$ is changed according
to a protocol up to a final time ${\tau}$. Work done during this
process is defined as \footnote{For a discussion of various
definitions of work, their relationship to each other and how that
affects the content and context of the fluctuation theorems see
\cite{Jarzynski2007}.}
\begin{equation}
\label{def:work0}
W=\int_{0}^{{\tau}} \rmd t \frac{\partial H}{\partial \lambda}
\dot{\lambda}(t),
\end{equation}
where an overdot denotes derivative with respect to time. Although the
Hamiltonian dynamics of the system is entirely deterministic, due to
the probabilistic nature of the initial conditions that are sampled
from the thermal phase space density, work is described by a probability
distribution  $\cP(W)$.     Note that thermal equilibrium is only assumed at
$t=0$ as part of the preparation. In general the evolved system is not in
thermal equilibrium. Jarzynski equality is the statement:
\begin{eqnarray}
\label{eq:Jarz}
\left \langle e^{-\beta W}\right\rangle \equiv \int \rmd W \cP(W) e^{-\beta
W}=e^{-\beta \Delta
F}.
\end{eqnarray}
Here $\Delta F\equiv F_\tau-F_0$, where $F_0$ and $F_\tau$ are free energies of
the system at thermal equilibrium
at inverse temperature $\beta$  with the parameter $\lambda$ assuming
values $\lambda_0$ and $\lambda_\tau$ respectively. Eq.(\ref{eq:Jarz})
is remarkable in that it relates an average of a thermodynamical
quantity over nonequilibrium processes to strictly equilibrium properties.

For the statement of Crooks's fluctuation theorem one defines the reverse
process in which the system is prepared initially at $t=0$ in the thermal state
$ \exp(-\beta H(\lambda_{\tau}))/Z_{\tau}$ and the parameter $\lambda$ is
changed in a time-reversed manner to assume the value $\lambda_0$ at ${\tau}$.
The probability distribution of work associated with this process is denoted by
$\cP_R(W)$. Crooks's fluctuation theorem states:
\begin{eqnarray}
\label{eq:Crooks}
\frac{\cP_F(W)}{\cP_R(-W)}=\rme^{\beta(W-\Delta F)},
\end{eqnarray}
where the subscript $F$ stands for forward process and $\Delta F$ is
defined as before. The Jarzynski equality follows from Eq.(\ref{eq:Crooks}) by
multiplying both sides by $\cP_R(-W)e^{-\beta W}$ and integrating over $W$.

\subsubsection{Quantum FTs}

The main difficulty in formulating fluctuation theorems for quantum mechanics
is defining work. Except for closed systems there is no agreement on a
definition of work in quantum mechanics. For closed systems there is general
agreement on the following  operational definition:
1) Measure the energy of the system using the Hamiltonian initially at
   $t=0$ to be $E^0_n$, thus `collapsing the wavefunction' to one of
   the eigenfunctions of the Hamiltonian at the initial time:
   $H(0)|\psi^0_n \rangle=E^0_n|\psi^0_n \rangle$. 2) Let the system evolve
under
   the time dependent Hamiltonian according to the prescribed
   protocol. 3) At the end of the protocol measure the energy of the
   system using the Hamiltonian at $t={\tau}$ to be $E_m^{\tau}$ thus
   collapsing the wavefunction to an eigenfunction of the Hamitonian
   at $\tau$: $H(\tau)|\psi^{\tau}_m\rangle=E^{\tau}_m|\psi^{\tau}_m\rangle$.
Work
   for this specified realization is defined as $W=E_m^{\tau}-E_n^0$.
   Since the system is closed one can interpret the change in energy
   of the system as work performed on the system. In classical
   mechanics of isolated systems work acquires a probabilistic feature
   only due to the sampling of the initial conditions,  since the
   dynamics is deterministic. In quantum mechanics work acquires an
   additional probabilistic feature from the dynamics: 
\begin{eqnarray}
\label{eq:QP(W)}
\cP(W)=\frac{1}{2\pi}\int \rmd u \  \rme^{-\imath u W}\tr[\rme^{\imath
u \hat{H}_H(\tau)}\rme^{-\imath u \hat{H}_H(0)}\hat{\rho}_\beta],
\end{eqnarray}
where the subscript H indicates Heisenberg operators. Jarzynski equality and
Crooks's fluctuation theorem can be proven in a few lines for a closed system
with this definition of work.

In this paper we develop an alternative approach based on a
microscopic model using the  open quantum systems paradigm. First
consider a classical harmonic oscillator, without a bath. Initial
position and momentum of the oscillator are sampled from the thermal
phase space density. The rest of the trajectory is entirely determined
by the  protocol of how the external force is applied. Work is
calculated using this deterministic trajectory according to
eq.(\ref{def:work0}). However, deterministic trajectory is strictly a classical
notion and cannot be applied to a general quantum mechanical system. A state
that is sampled from the thermal density matrix in general does not have a well
defined position and momentum. Furthermore the time evolution usually causes
the wavefunction to spread further.  We cannot talk about the quantum
oscillator being at one point in space having a certain velocity and moving in
a deterministic continuous trajectory as a function of time.

Next consider the same classical model with a heat bath. For each realization
of the protocol, the initial data for both the system oscillator and the bath
are sampled from the initial phase space density. The initial data for the bath
determines the noise for that particular realization. The system oscillator
follows a trajectory determined by a combined action of the deterministic force
$f(t)$ and the stochastic force $\xi(t)$. Although the noise is stochastic,
each realization of the experiment corresponds to a unique noise and hence a
unique trajectory. The definition of work in terms of trajectories is
unaffected.

It is a simple yet subtle and deep point how the interaction with a bath would
help to define a trajectory for a quantum particle. To understand this
conceptually we adopt the decoherent or consistent histories viewpoint of
quantum mechanics as described below.

\subsection{Trajectories in quantum mechanics}

Trajectories which are well defined in classical mechanics are generally ill
defined in quantum mechanics
except under certain conditions. We shall spell out these conditions
here. Let us begin with something simple, such as a quantum particle in
motion. 

In a closed quantum system S, namely, a system subjected to no outside
(environmental) influence except for its own quantum fluctuations, the closest
entity to its “trajectory” is a wave packet moving
with a certain group
velocity but that also spreads in time due to the Heisenberg uncertainty
relation between the variance of the canonical variables, position and momentum
in this case.
The same system at a finite temperature is no longer
closed
because for it to exist at finite temperature it must be or have been
in contact with a source with energy exchange or a bath B kept at
non-zero temperature. The influence of the environment E (we call an
E a bath B if it is described by a thermal density matrix with
inverse temperature $\beta$) has complicated and interesting
consequences. This is the subject of open quantum systems.

There are at least two major effects an environment brings  in
`opening up' a closed quantum system: a) it turns the original
Hamiltonian (unitary) dynamics to dissipative (nonunitary) dynamics
-- this refers to energy flow from the system to the environment, b)
fluctuations in the environment decoheres the quantum system -- this
refers to quantum phases of the system being dispersed into the
environmental variables.  The latter is responsible for  shaping the
notion of trajectories in quantum system and there are precise
conditions pertaining to the features  of the environment (e.g., ohmic
spectral density, high temperature) whereby it becomes physically
well-defined in a measurement. One way is to construct the reduced
density matrix of the open quantum system and look at whether and how
quickly its off-diagonal elements decay in time, leaving the
system's statistical state describable by an approximately diagonal density
matrix with respect to
some physically meaningful basis (related to measurement instruments
and interaction, such as Zurek's `pointer basis' \cite{Zurek1982}).
This time, called decoherence time, marks the appearance of classical
features, because after it is effectively decohered this open system
is adequately described by probabilities rather than amplitudes, its
quantum phase information is lost (more accurately, dispersed into or
shared by the multitude of environmental degrees of freedom). This
process is captured by the stochastic equations, the most common
forms are the master equation, the Langevin and the Fokker-Planck
equations.

What distinguishes these equations is the presence of noise or fluctuations in
the environment, and dissipative dynamics of the open (reduced) system,
depicting the two distinct features of open system dynamics.  In general two
kinds of noise exist in any quantum system,  the intrinsic quantum noise
entering in the Heisenberg relation which exists for
all systems even at zero temperature, and thermal noise from a finite
temperature bath. Both contribute to decohering a quantum system
although the thermal noise usually overwhelms.

There are many ways to characterize a quantum system as approaching its
classical limit. The
familiar cases are the correspondence principle, the Bohr-Sommerfeld
rules in quantum mechanics, the description of Maxwell-Boltzmann
distribution as limits of the Fermi-Dirac and the Bose-Einstein
distributions, or the more simplistic $h \rightarrow 0$ or `at high
temperature' stipulations. One can show that the coherent state is
the `most classical' of quantum states \cite{Zurek1993}. One can derive
an uncertainty function at finite temperature \cite{Hu1995}
or
equivalently calculate the entropy function and be able to demarcate
the transition from the quantum noise- dominated regime to the
thermal noise- dominated regime. There is significant advance in the
last two decades in our understanding of the  quantum classical
correspondence. (See e.g., \cite{Feng1997}) Decoherence is at the
heart of the quantum to classical transition issue, and the main
cause of it is noise of all forms, either in the fluctuations of the
environment, or in the separation of neighboring coarse-grained histories, and
in
the accuracy of the measurement devices and procedures. We will use the
decoherent or
consistent history \cite{Omnes1992,Gell-Mann1993,Griffiths1984} viewpoint for
conceptual
clarity, especially pertaining to the issue of trajectories but adopt
the environment-induced decoherence ({\it envdec}) scheme for computations,
as it is technically easier to manipulate.

\subsubsection{Decoherence Functional in {\it Dechis} and Influence Functional
in {\it Envdec}}

The main idea of {\it dechis} approach is to define a history $\alpha$ by a set
of
projection operators $P_\alpha(t_k)$ acting at times $t
_k$. As a special case we consider projections in position basis.
These kind of histories are naturally implemented in the path integral
approach. The projectors are represented by window functions
$w_\alpha\left[ x(t_k) \right]$, which take on unit value if the
instantaneous configuration satisfies the requirement of the history
$\alpha$, and vanish otherwise. As a limiting case we mention a fine-grained
history, for which the path is specified exactly at all times and is
assigned an amplitude $\exp(\imath S/\hbar)$ as usual. It is useful to define
the decoherence
functional of two histories $\alpha$ and $\beta$ by \cite{Gell-Mann1990}:
\be
\label{eq:decfun}
\mathcal{D}[\alpha,\beta]=\int Dx Dx'
e^{i(S[x]-S[x'])/\hbar}\rho(x(t_i),x'(t_i);t_i)\left\{ \prod_k
w_\alpha[x(t_k)]
\right\}\left\{ \prod_l
w_\beta[x'(t_l)]
\right\}.
\ee 
The product over $k$ and $l$ can be discrete or continuous as is the case in
Section (\ref{sec:decfun}).
The probability of a given history $\alpha$ is given by the diagonal element of
the
decoherence functional: $P[\alpha]=\mathcal{D}[\alpha,\alpha]$. For
classical trajectories it is required that the probability of a coarse
grained history to be the sum of its constituents. For an arbitrary
set of histories quantum
interference effects lead to a violation of the probability sum rule:
$P[\alpha\vee\beta]=\mathcal{D}[\alpha,\alpha]+\mathcal{D}[\beta,\beta]+2\rm{Re
}\mathcal{D}[\alpha,\beta]\neq
P[\alpha]+P[\beta]$. If a set of histories can be identified for which
the real part of the off-diagonal elements of the decoherence
functional vanishes (or are much smaller than the diagonal elements for
approximate decoherence), probabilities can be assigned to individual
histories. The challenge is to identify the conditions under which,
and to what extent, the decoherence condition is satisfied.

Technically the environment-induced decoherence ({\it envdec}) program
is easier to implement, the relation between these two programs are
explained or illustrated in \cite{PazZurek1993,Dowker1992}. This is what we
will do by way of the QBM model presented in the next section. We will argue
that for histories obtained by coarse graining the environment sufficiently,
and the system of interest to some extent (determined by the strength of
noise),
an  approximate decoherence condition can be satisfied to a specified degree of
accuracy.   At the other end,
if quantum interference between particle histories continues to play
an overbearing role, decoherence is not consummated,  the classical
world is not reached and the concept of trajectories is ill-defined.

The quantum open system formulation, via the influence functional, provides one
with a clear
perspective in the organic relation between the processes of
fluctuations / noise, correlation, decoherence and dissipation and how
they enter in the transition from the quantum to the classical world
with the intermediate stochastic and semiclassical regimes.
While it is useful to explain this with the aid of stochastic
equations which we will derive below, the key idea can be put
succinctly: The stronger the effect of noise in the environment the
more efficient it decoheres the quantum system and the clearer the
classical notion of trajectory can be defined and used for the
description of a quantum particle. The important new understanding is the
existence of a stochastic regime between the quantum and the
classical, and how quantum features are expressed in terms of
classical stochastic variables \footnote{A famous case is the transcription of
Gaussian quantum fluctuations in the environment as classical noise
via the Feynman-Vernon identity \cite{FeynmanVernon1963}.}.

\subsubsection{Worldline Influence Functional Formalism}

Thus far we learned that the decoherence of a quantum system due to
the noise arising from a coarse-grained environment is instrumental to
the emergence of a classical world. How strongly the system is
coupled to its environment(s), the nature of the noise from the
environment and its temperature all enter in determining how completely
the system is decohered, and there is always a stochastic component in
the open system's dynamics governed by a Langevin equation or its
(near) equivalent master or Fokker-Planck equations. Almost complete
decoherence is a
necessary condition for a classical description which in this context,
is what trajectories are predicated upon. Under this condition a
powerful  approach called  the worldline (WL) influence functional
(IF) formalism has been used effectively for more than two decades in
nuclear / particle physics communities, see e.g., \cite{Schubert2001}. We 
shall only mention its key features so as to bring out its relevance to the
present problem but skip all the details.

The influence functional technique of Feynman and Vernon
\cite{FeynmanVernon1963}, or the closely related closed-time-path
effective action method of Schwinger \cite{Schwinger1961} and Keldysh
\cite{Keldysh1964} are initial value (in-in) formulations which are
particularly suitable for exploring the time evolution of many body systems,
unlike the S-matrix (in-out) formulation used for calculating scattering
processes.  In general this yields a  nonlocal and nonlinear coarse-grained
effective action (CGEA) for the system's motion. The CGEA may be used to treat
the nonequilibrium quantum dynamics of interacting particles.  Take for example
the QBM model:  When the particle trajectory becomes largely well defined as a
result of effective decoherence due to interactions with the environment, with
some degree of stochasticity caused by noise,  the CGEA can be meaningfully
transcribed into a stochastic effective action, describing stochastic particle
motion. The evolution propagator for the reduced density matrix of the open
system is dominated by
the particle trajectory giving the extremal solution of the real part of the
CGEA.
Stochastic fluctuations around the decohered semiclassical
trajectories are described by the imaginary part of the CGEA.  For
further technical details, see \cite{JohnsonHuArxiv,Johnson2002}.

When the back-action of the environment is taken into account the dynamics of
the open system will in general be non-Markovian as it contains memories, and
the noise in the environment is generally colored, as it contains many time
scales characterized by its spectral density and vary with temperature.
Dissipation in the open system dynamics is controlled and balanced by the noise
in the environment as manifested in the existence of fluctuation-dissipation
relations between these two sectors. What is more important, because the
influence action includes the back-action of the environment in a
self-consistent manner,  the worldline is not merely a prescribed classical
entity, or a simple solution to an equation of motion at the tree level (in
truth, with an ever-present stochastic component), but rather, a dynamical one,
as the result of constant negotiation between the open system and its
environments at all times. This is the special beauty of the  IF method.

\section{The QBM model}

In this section we describe the salient features of the bilinear QBM model with
a general environment following \cite{Hu1992} for non-Markovian dynamics and
write down the solutions of the Langevin equation following
\cite{Calzetta2003}. We focus on the stochastic dynamics of the quantum open
system, which incorporates the effects of the environment. This  will play a
crucial role in our formulation of FTs in the following section.

A closed quantum system can be partitioned into several subsystems according to
the relevant physical scales. If one is interested in the details of one such
subsystem, call it the
distinguished, or relevant system, which interacts with the other
subsystems comprising the environment, the details of which are not of
interest, one can coarse-grain the information in the environment but
keep its overall influence on the distinguished subsystem of interest,
thereby rendering it an open system.  This influence is best
captured by the influence functional technique of Feynman and Vernon
\cite{FeynmanVernon1963} which we use here. Let us assume for simplicity the
system S is comprised of a simple harmonic oscillator with position and
momentum $(x,p)$ linearly
coupled to a heat bath B consisting of N harmonic oscillators with positions
and momenta $(q_n, p_n)$ with $n=1,..N$ and allow the system to be driven by a
time-dependent external force $f(t)$. The Hamiltonian describing the dynamics
of the combined system is given by:
\be
\label{eq:H_T}
H_{T} = (H_S) + [H_{B}+H_I+H_R],
\ee
where
\begin{eqnarray} 
  (H_S) &=& \frac{p^2}{2M}+\frac{1}{2} M \Omega^2 x^2-f(t)x,  \\ \nn
  [H_{B}+H_I+H_R]  &=& \sum_{n=1}^N  \left[ \frac{p_n^2}{2
  m_n}+\frac{1}{2} m_n
\omega_n^2\left(q_n-\frac{c_n}{m_n \omega_n^2}x\right)^2\right].
\end{eqnarray}
The system oscillator is coupled to the bath via the  linear
interaction term $H_I=(\sum_n c_n q_n) x$. A renormalization of the
potential via $H_R=\sum_n\frac{c_n^2}{2 m_n\omega_n^2}x^2$ preserves the
physical (observed) frequency of the system oscillator for any
system-bath coupling as will be shown below. Without the
renormalization term the potential might have no minimum  and the
thermal state could not be defined. The remainder in the square
bracketed quantity in the above equation is $H_B$. This is the QBM
model.

The QBM model is used extensively in the open quantum systems
literature thanks to its exact solubility and  its generality. The
solubility is due to the linearity of the model. The generality may
not be immediately obvious. Representing the environment by a set of
simple harmonic oscillators might appear to be a serious restriction
to weak influences on the system, because of its linearity. An argument for the
generality of the
model is given by Caldeira and  Leggett \footnote{``For most cases of
interest, at least when the system variable is macroscopic, this
assumption is physically reasonable; in that case the environment is
usually also (geometrically) macroscopic and the interaction of the
system with any one environmental degree of freedom is generally
proportional to the inverse of the volume, while the characteristic
energy of such a degree is of freedom is
volume-independent."\cite{Caldeira1983}}.
The applicability of the model is limited to cases where the influence of the
system on each bath mode is weak. This does not imply that the influence of the
bath as a whole on the system is weak as well.
 The Brownian particle interacts with a very large number of environmental
degrees of freedom. The effect of these interactions can add up to yield strong
dissipation, fluctuations and decoherence for the Brownian particle.

The combined system being closed, the Hamiltonian $H_T$ gives unitary
evolution, its density operator $\rho$ obeys the von-Neumann equation
\be
\imath \hbar \frac{\partial \rho(t)}{\partial t}=[H_T,\rho(t)].
\ee
An alternative description closer in spirit to (but should by no means be
identified with) that of a trajectory in phase space which gives a full and
equivalent description is by way of the Wigner function \cite{Hillary1984}. For
a closed system
such as S by itself, without any interaction with its  environment,
$\cW(X,p,t)$ is defined with the new variables  $X\equiv(x+x')/2$ and $y\equiv
x'-x$:
\begin{equation}
\label{eq:wigner}
\cW(X,p,t)=\frac{1}{2\pi \hbar} \int_{ }^{ } dy
e^{\frac{\imath}{\hbar}p y}\rho(X-y/2,X+y/2,t).
\end{equation}
Because of its appearance in phase space variables it is often said
that the Wigner function is the quantum correspondence of the
classical phase space density \cite{Habib1990} and the peak of the
Wigner function coincides with the classical trajectory in phase
space. This is an erroneous statement. Wigner function gives as complete a
description as that provided by the density matrix  because
it contains full quantum phase information. As such it can take on
negative values \footnote{Under special conditions for Gaussian
systems such as a free simple harmonic oscillator (closed system)  or
one which interacts bilinearly  with an ohmic bath at high temperature
(an open system) the Wigner function is positive definite for all
times. The quantum and classical dynamics have the same form in the
equations of motion \cite{Habib2004}.  For more general conditions by including
environmental influence the reduced Wigner function (defined later) may become
positive definite at late times after the system has  sufficiently been
decohered. This can indeed be used as a criterion for the appearance of
classicality and the condition for the trajectory notion to be safely adopted
in a quantum open system.}.

However what we are interested in is how the system S behaves under the
influence of its environment, in this case a heat bath B at temperature
$1/\beta$.  The state of the open system at any one time is completely
specified by the reduced density matrix $\rho_r$, which is obtained from the
density matrix of the combined system by integrating out the bath degrees of
freedom. In position representation it is given by:
\begin{equation}
\rho_r(x,x',t)=\int \prod_n d q_{n} \rho(x,\{q_{n}\},x',\{q_{n}\},t).
\end{equation}

Because it incorporates the back-action of the environment the time evolution
of the reduced density matrix of the open system is nonunitary and in general
non-Markovian. The reduced density matrix of the system oscillator at ${t_f}$
can be obtained from the reduced density matrix at some earlier time $t_i$ via
\cite{Grabert1988}:
\begin{equation}
\label{eq:propagation}
\rho_r(x_f,x'_f,{t_f})= \int d x_i d x'_i
J(x_f,x'_f,{t_f};x_i,x'_i,t_i) \rho_r(x_i,x'_i,t_i),
\end{equation}
where $J$ is the propagator.
If the system and the bath are initially uncorrelated and the bath is in a
Gaussian state the propagator $J$ can be calculated exactly:
\begin{equation}
\label{eq:propagator}
J(x_f,x'_f,{t_f};x_i,x'_i,t_i)=\int_{x(t_i)=x_i}^{x({t_f})=x_f}Dx\int_{
x'(t_i)=x'_i}^{x'({t_f})=x'_f}Dx'
e^{\frac{\imath}{\hbar}\left( S_S[x]-S_S[x']+S_{IF}[x,x']  \right) }.
\end{equation}
Now introduce the following notation: for functions $A(s), B(s)$ and kernel
$K(s,s')$ define
\be
\label{def:dot}
A \cdot K \cdot B\equiv \int_{t_i}^{{t_f}}ds \int_{t_i}^{{t_f}}ds'
A(s) K(s,s') B(s').
\ee
In terms of the new variables the exponent appearing in
Eq.(\ref{eq:propagator}) can be written as:
\begin{eqnarray}
\label{eq:S-S}
 S_S[x]-S_S[x']&=&-M\dot{X}({t_f})y_f+M\dot{X}(t_i) y _i+ y  \cdot L_0
 \cdot X,  \\
 \label{eq:SIF}
 S_{IF}[x,x'] &=& -  y  \cdot \mu \cdot X+\frac{\imath}{2}  y  \cdot
 \nu \cdot  y,
\end{eqnarray}
where $L_0(t,t')=M(\frac{d^2}{dt^2}+\Omega^2)\delta(t-t')$. The
kernels $\mu(s,s')$ and $\nu(s,s')$ are called the dissipation and
noise kernels, respectively. For the special case, when the heat bath
is in a thermal state of the bath Hamiltonian $H_B=\sum_{n=1}^N
\left[ \frac{p_n^2}{2 m_n}+\frac{1}{2} m_n \omega_n^2x_n^2\right]$, these
kernels are given by:
\begin{eqnarray}
\label{eq:Dissipation}
\mu(t,t') &=&
\sum_{n=1}^{N}\frac{c_n^2}{m_n\omega_n}\sin[\omega_n(t-t')]\Theta(t-t'),    \\
\label{eq:Noise}
\nu(s,s') &=&
\sum_{n=1}^{N}\frac{c_n^2}{2m_n\omega_n}\coth(\frac{\beta \hbar
\omega_n}{2})\cos[\omega_n(t-t')].
\end{eqnarray}

In the equivalent description in terms of the Wigner function one
defines a reduced Wigner function $\cW_r$ in terms of the reduced
density matrix formally in the same way as in Eq. (\ref{eq:wigner})
(denoted by a subscript $r$). Using eqs.(\ref{eq:wigner}-\ref{eq:propagator})
it  can be show that the reduced Wigner function evolves from time $t_i$ to a
later time ${t_f}$ via
\begin{eqnarray}
\nonumber
 \cW_r(X_f,p_f,{t_f})=\frac{1}{2\pi \hbar} \int_{}^{} d y _f
e^{\frac{\imath}{\hbar}p_f  y _f}\int d x_i d x'_i
\int_{x(t_i)=x_i}^{x({t_f})=X_f- y _f/2}Dx\int_{x'(t_i)=x'_i}^{x'({t_f})=X_f+ y
_f/2}Dx'\\
 \times  e^{\frac{\imath}{\hbar}\left( S_S[x]-S_S[x']+S_{IF}[x,x']  \right) }
\int_{}^{} dp_i e^{-\frac{\imath}{\hbar}p_i
(x'_i-x_i)}\cW_r(\frac{x_i+x'_i}{2},p_i,t_i).
\end{eqnarray}
First we perform a functional change of variables from the variables
$x(t),x'(t)$ to $X(t)=(x'(t)+x(t))/2,\ y(t)=x'(t)-x(t) $. We also perform a
regular change of variables from $x_i, x'_i$ to $X_i=(x'_i-x_i)/2,\ y
_i=x_i'-x_i$. The Jacobian determinant for both change of variables is one. 
Then we use eqs.(\ref{eq:S-S},\ref{eq:SIF}) and define $L=L_0-\mu$ to obtain:
\begin{eqnarray}
\nonumber
\cW_r(X_f,p_f,{t_f})=\int d X_i\int_{}^{} dp_i \cW_r(X_i,p_i,t_i) \frac{1}{2\pi
\hbar}  \int d y _f e^{\frac{\imath}{\hbar}p_f  y _f}\int d  y _i
e^{-\frac{\imath}{\hbar}p_i  y_i} \\
\times \int_{ y(t_i)= y_i}^{ y({t_f})= y_f}D
y\int_{X(t_i)=X_i}^{X({t_f})=X_f}DX
e^{\frac{\imath}{\hbar}\left(-M\dot{X}({t_f}) y_f+M \dot{X}(t_i) y_i +
y\cdot L\cdot X-\frac{1}{2} y\cdot \nu \cdot  y  \right)}.
\end{eqnarray}
The functional integral over $ y$ is Gaussian and can be evaluated formally to
give:
\begin{equation}
\int_{ y(t_i)= y_i}^{ y({t_f})= y_f} D y e^{\frac{\imath}{\hbar}(
y\cdot L\cdot X-\frac{1}{2} y\cdot \nu \cdot  y)}
=\sqrt{\frac{1}{det(\nu/2\pi\hbar)}}e^{-\frac{1}{2\hbar}(L\cdot
X)^T\cdot \nu^{-1}\cdot(L\cdot X)}.
\end{equation}
For the type of noise kernels displayed in Eq.(\ref{eq:Noise}) the outcome of
this functional integral is independent of the endpoints $ y_i$ and $ y_f$,
irrespective of the distribution of bath frequencies. As a result the integral
over $ y_i$ and $ y_f$ is trivial and gives $(2\pi
\hbar)^2\delta(M\dot{X}(t_i)-p_i)\delta(M\dot{X}({t_f})-p_f)$. We have
\begin{eqnarray}
\nonumber
\cW_r(X_f,p_f,{t_f})=\frac{2\pi \hbar}{\sqrt{det(\nu/2\pi \hbar)}} \int d
X_i\int_{}^{} dp_i \cW_r(X_i,p_i,t_i)  \\
\times \int_{X(t_i)=X_i}^{X({t_f})=X_f}DX (2\pi
\hbar)^2\delta(M\dot{X}(t_i)-p_i)\delta(M\dot{X}({t_f})-p_f)
e^{-\frac{1}{2}(L\cdot X)^T\cdot (\hbar \nu)^{-1}\cdot(L\cdot X)  }.
\end{eqnarray}
Next we do another functional change from $X(t)$ to $\xi(t)$ where
\begin{equation}
\label{eq:Change}
 X(t) \rightarrow \{X_i=X(t_i),\ p_i=M\dot{X}(t_i),\ \xi(t)=(L\cdot
 X)(t)\}.
\end{equation}
For linear change of variables the Jacobian functional determinant is
independent of $\xi$.
To ensure that the boundary condition at ${t_f}$ is satisfied we need to place
a delta function inside the new path integral. The net effect of the functional
change of variables is:
\be
\int_{X(t_i)=X_i}^{X({t_f})=X_f}DX\delta(M\dot{X}(t_i)-p_i)
\rightarrow  \int D\xi \delta(X_\xi({t_f})-X_f),
\ee
where $X_\xi(t)$ is the solution of the Langevin equation $(L\cdot X_\xi)(t) =
\xi(t)$ with the initial conditions $(X_i,p_i)$.
 After this functional change we obtain:
\begin{eqnarray}
\nonumber
 \cW_r(X_f,p_f,{t_f})&=& \int d X_i\int_{}^{} dp_i \cW_r(X_i,p_i,t_i)\\
 \nonumber
&&\quad \times \int \frac{D\xi}{\sqrt{det(2 \pi \hbar \nu)}}
\rme^{-\frac{1}{2}\xi^T\cdot (\hbar \nu)^{-1}\cdot\xi}
\delta(M\dot{X}_\xi({t_f})-p_f) \delta(X_\xi({t_f})-X_f)\\
\nonumber
&=&  \int d X_i\int_{}^{} dp_i \cW_r(X_i,p_i,t_i)\int D\xi P[\xi]
\delta(M\dot{X}_\xi({t_f})-p_f) \delta(X_\xi({t_f})-X_f)\\
\label{eq:<>}
&=& \left \langle \delta(M\dot{X}_\xi({t_f})-p_f) \delta(X_\xi({t_f})-X_f)
\right\rangle _{(X_i,p_i),\xi}.
\end{eqnarray}
Here $\xi(t)$ is a random noise with Gaussian statistics and is characterized
by its mean and variance:
\begin{eqnarray}
\left \langle \xi(t)\right\rangle &=&0. \\
  \label{eq:noise}
\left \langle \xi(t)\xi(t')\right\rangle &=&\hbar \nu(t,t').
\end{eqnarray}
Furthermore since the system and bath are assumed to be uncorrelated initially:
\begin{equation}
\left \langle X_i \xi(t)\right\rangle =\left \langle p_i \xi(t)\right\rangle
=0.
\end{equation}
Eq.(\ref{eq:<>}) has a clear interpretation. The dynamics of the reduced Wigner
function is identical to the dynamics of  the phase space density of a
stochastic classical system described by the Langevin equation $(L\cdot
X_\xi)(t) = \xi(t)$.

As argued in \cite{Calzetta2003} the Langevin equation provides a more
detailed description of the dynamics than the master equation, in the
sense that the class of quantum correlation functions which may be
retrieved from the Langevin equation is larger than the corresponding
class for the master or Fokker-Planck equations unless the dynamics is
Markovian. Work as defined in Eq.(\ref{eq:def_work}) is an example of
this kind of quantity, since its statistics requires the calculation
of multi-time correlations.

It is important to realize that this method gives exact quantum mechanical
results at any parameter regime, including arbitrarily low temperatures. The
fact that solutions $X_\xi$ of a classical Langevin equation are used in
eq.(\ref{eq:<>}) should not be conjured as having made a semiclassical
approximation as was done in e.g. \cite{Chernyak2004}.

The effect of environment-induced decoherence at work which validates
the notion of a physical trajectory is implicitly contained in this method
(depending on the temperature of the bath and its spectral density), not
extrinsically introduced by hand. Since these processes are dynamically and
self-consistently determined  no semiclassical approximation has been made
specifically in the derivation. The real challenge is in the interpretation of
the physical variables in light of  quantum measurement theory, as we discussed
previously. In the following section we discuss under what conditions physical
trajectories emerge from the dechis formalism.

\subsection{Decoherence Functional}
\label{sec:decfun}

We consider histories where the system variable $X$ is specified to
follow a trajectory $\chi(t)$ with a given accuracy $\sigma(t)$, while
the environment variables are left completely unspecified. For
technical reasons it is convenient to use Gaussian, rather than sharp,
window functions $w_\chi[x(\cdot)]$. In the path integral this roughly
corresponds to using $\exp\left\{ -\int dt \frac{(x(t)-\chi(t))^2}{2
\sigma^2(t)}
 \right\} $ in eq.(\ref{eq:decfun}).
Furthermore we introduce window functions at every instance of time rather than
at discrete time intervals.  The set of Gaussian window functions with this
property acts as a noise term in the influence action. This can be seen in
eq.(\ref{eq:expdecfun}) where the noise kernel always occurs in the combination
$\nu+(2 \sigma^2)^{-1}$.
There is some error introduced due to the overlap of projectors
defined as above. As a result we will be talking
about approximate decoherence. In addressing the diagonal and
off-diagonal elements of the decoherence functional it is convenient
to define $U=(\chi'+\chi)/{2}$ and $u=\chi'-\chi$.
In \cite{Calzetta2008} it is shown that
the decoherence functional for two histories $\chi(t)$ and $\chi'(t)$ defined
via these
projectors is approximately given by:
\be
\label{eq:expdecfun}
|\mathcal{D}[U-u/2,U+u/2]|\sim \exp\left\{ -\frac{1}{2}(L\cdot U)^T\cdot
(\nu+(2\sigma^2)^{-1})^{-1}\cdot(L\cdot
U)-\frac{1}{2}u\cdot(\nu^{-1}+2\sigma^2)^{-1}\cdot u \right\}.
\ee
Here we again used a compact notation as before and $\nu$ is defined in
eq.(\ref{eq:noise}). The off-diagonal
elements vanish as a Gaussian for $u \gtrsim \sqrt{\nu^{-1}+2\sigma^2}$.
Hence an approximately consistent set of histories can be obtained by
picking histories that differ by at least this amount. However if the Langevin
noise is weak such that $\nu^{-1}\gg \sigma^2$, we run into trouble because now
decoherence condition requires $u\gtrsim \sqrt{\nu^{-1}}\gg
\sigma$. Histories of accuracy $\sigma$ in a set do not interfere with each
other only if they are separated by a distance much larger than
$\sigma$. This suggests that we cannot account for all probabilities within
such a set.

We conclude that the accuracy should be adjusted to the noise level
by: $\sigma^2\sim \nu^{-1}$. Then the decoherence condition requires
that $u\gtrsim \sigma$. Now we can have a set of histories which
decohere approximately and for which the resulting probabilities add up to one.
``A picture of the system
evolution based on actual nearly classical trajectories may only
result from a compromise whereby the accuracy of observations is
adjusted to the noise level, $\sigma^2\sim \nu^{-1}$ where $\s$ is the
accuracy at which the trajectories are defined.
Larger noise for a given $\sigma$ means more decoherence
but less predictability; for a weaker noise, predictability is only
limited by the Heisenberg bounds, but individual trajectories will
not decohere.\footnote{We will continue the exploration of this regime
in a sequel paper.} If we are satisfied with predictability within the
limits imposed by the Langevin equation, then in the strong noise
limit we may consider individual trajectories as depicting physical
reality." \footnote{Quotation is from \cite{Calzetta2008}, p. 89.}  This
condition is ordinarily satisfied at high enough temperatures where the quantum
and classical trajectories agree, even
for non-Markovian dynamics, as we will see below.

For a given accuracy $\sigma$ the higher the temperature the stronger the noise
and the more effective it brings about decohering histories into trajectories.
Thus even at moderate temperatures and for relatively weak noise by judicious
choices of the coarse graining measure $\sigma$ decoherence can be effective
enough to warrant the notion of trajectories. It is in this regime where
deviations from FTs can be identified using this method. At low enough
temperatures when no reasonable set of histories decohere and the notion of
trajectories lacking, we cannot say the FTs are violated (even though it
appears reasonable to doubt its validity) because the contents of FTs may be
phrased  without invoking trajectories.   For completeness of technical
presentation we provide a low temperature expansion in section
(\ref{sec:lowtemp}).

\section{Solutions of the Langevin Equation}

It is convenient to rewrite the Langevin equation as:
\begin{eqnarray}
\label{eq:Lan1}
 M\ddot{X_\xi}(t)&+&2M\int_{t_i}^{t}\rmd s
 \g(t-s)\dot{X}_\xi(s)+M\Omega^2 X_\xi(t)=f(t)-2M\g(t-t_i)X_i+\xi(t),\\
 \g(t-s)&=&\frac{1}{M}\sum_{n=1}^N\frac{c_n^2}{2m_n\omega_n^2}\cos[
\omega_n(t-s)],
\end{eqnarray}
where $\gamma$ is the damping kernel defined as the antiderivative of the
dissipation kernel $\mu$ \cite{Hu1992}. It is related to the noise kernel by
the fluctuation-dissipation relation (\ref{eq:QFDR}) (See
\cite{FlemingAnnPhys,FlemingThesis} for further exposition of the meaning and
respective roles of $\nu$ and $\gamma$.)

where $\gamma$ is the damping kernel defined as the anti-derivative of
the noise kernel $\nu$ \cite{Hu1992}. (See
\cite{FlemingAnnPhys,FlemingThesis} for further exposition of the meaning and
respective roles of $\nu$ and $\gamma$.) It is related to the
noise kernel by the fluctuation- dissipation relation (\ref{eq:QFDR}).
In the rest of the paper we will drop the subscript $\xi$. The
Langevin equation (\ref{eq:Lan1}) is a linear integro-differential
equation. The effect of interactions of the system oscillator with the
bath is contained in a nonlocal  potential and renormalization of the
potential. One can think of the nonlocal potential as the system
oscillator interacting with its own past, where the interaction is
mediated by the bath. A formal solution to this equation can be
obtained in terms of the homogenous solutions to the LHS of Eq.(\ref{eq:Lan1})
with $t_i$ set equal to zero. Let us call the two linearly independent 
homogenous solutions $h(t)$ and $g(t)$ such that:
 \begin{eqnarray}
 h(0)=\dot{g}(0)=1;\qquad \dot{h}(0)=g(0)=0.
 \end{eqnarray}
 The formal solution of the Langevin equation is then:
 \begin{equation}
 \label{eq:LanSol}
X(t)=X(t_i)h(t-t_i)+p(t_i)g(t-t_i)+\int_{t_i}^t\rmd t' g(t-t')\left[
f(t')+\xi(t')-2MX(t_i)\g(t'-t_i)\right].
 \end{equation}
 $h(t)$ and $g(t)$ can be calculated using the Laplace transforms:
\begin{equation}
\label{eq:HGLap}
\hat{h}(s)=\frac{2\hat{\g}(s)+s}{s^2+2 s\hat{\g}(s)+\Omega^2}, \qquad
\hat{g}(s)=\frac{1/M}{s^2+2s\hat{\g}(s)+\Omega^2}.
\end{equation}
where the hat indicates Laplace transform. These expressions show the relation
between the two linearly independent homogenous solutions:
\begin{eqnarray}
s\hat{h}(s)=1-M\Omega^2\hat{g}(s), \qquad
&&sM\hat{g}(s)=\hat{h}(s)-2M\hat{\g}(s)\hat{g}(s),\\
\label{eq:HGRel}
\dot{h}(t)=-M\Omega^2g(t), \qquad &&M\dot{g}(t)=h(t)-2M\int_0^t \rmd s
\g(t-s)g(s).
\end{eqnarray}

\subsection{Initial State Preparation}
\label{sec:prep}

The derivation of classical mechanical FTs for closed systems requires the
closed system to be in a thermal state. As pointed out earlier our derivation
of the Langevin equation (\ref{eq:Lan1}) assumes an uncorrelated initial state
in which the bath is in the thermal state of its own Hamiltonian $H_B$.
Such a state is obviously not the thermal state of the combined system 
and it is not stationary for any choice of the system's initial state. For this
reason the uncorrelated initial state is not appropriate for applications to
FTs. This observation is valid even for the classical Brownian motion model and
is
therefore not due to a quantum mechanical effect.

Assume that
the bath oscillator frequencies form a continuum. It is customary to define the
spectral density of the bath as
\begin{equation}
\label{eq:contlim}
J(\omega)\equiv \sum_n \frac{c_n^2}{2 m_n \omega_n}\delta(\omega-\omega_n)
\end{equation}
and interpret $J(\omega)$ as a continuous function. The resulting
Langevin dynamics is truly dissipative, in the sense that $\lim_{t\rightarrow
\infty}g(t),\g(t)=0$. Physically, true dissipation corresponds to a positive
average heat rate at all times. If the
spectrum of bath frequencies is discrete, the resulting damping kernel is
oscillatory. This is the case even for an infinite but countable number of
discrete frequencies. As a result after some (possibly very long) time there
may be average heat flow from the bath into the system. By true dissipation we
mean a definite arrow of time for all times.
Under these assumptions it can be shown that \cite{FlemingSubasi} if
the uncorrelated initial state is prepared at the infinite past, for
times $t > 0$ the dynamics of the system oscillator is indistinguishable
from that of a combined system + bath thermal state
preparation. In other words the effect of a thermal initial state can be
achieved by
allowing the uncorrelated system to thermalize for an infinite amount of time.
At t=0 the system density matrix is Gaussian. Means and variances of position
and momentum are equal to those of the combined
thermal state of QBM given in \cite{Weiss2008}: 
\begin{align}
 \sigma_{xx} &= \frac{1}{M \beta} \sum_{r=-\infty}^{\infty} \frac{1}{\Omega^2
+\nu_r^2 + 2 \nu_r \hat{\gamma}(n_r)}\\
 \sigma_{pp} &= \frac{M}{\beta} \sum_{r=-\infty}^{\infty} \frac{\Omega^2+2
\nu_r \hat{\gamma}(\nu_r)}{\Omega^2
+\nu_r^2 + 2 \nu_r \hat{\gamma}(n_r)}
\end{align}
where $\nu_r=2\pi r/\hbar \beta$ are the bosonic Matsubara frequencies.
These variances differ from those corresponding to a Boltzmann distribution
with
respect to the system Hamiltonian alone. The differences start at second order
in the coupling strength between the system and the bath. In the literature
ignoring these differences is sometimes referred to as the \textit{weak
coupling
approximation}. The results of this paper \textit{do not} depend upon the
\textit{weak coupling approximation} in this sense.

It is worth emphasizing that the equivalence of ensemble preparations is not
just on the level of reduced density matrices,
which can give only single-time correlations for general
non-Markovian dynamics. As pointed out before FTs require
multi-time correlations which can be obtained via the Langevin
equation. It is the equivalence of the trajectories for $t > 0$ that
can be shown exactly for the two preparations mentioned.
This means
that any quantum mechanical correlation function involving only the open
system variables and times larger than zero will be identical in both
preparations\,\cite{FlemingSubasi}.

As a result the trajectories we obtained in the previous section can be used to
describe a thermal state as long as we take $t_i\rightarrow -\infty$ and assume
a continuous spectrum for the bath frequencies. The thermal state preparation
procedure is as follows: at the infinite past the system and bath are in a
product state: the bath is in the thermal state, and the system is in an
arbitrary state. The combined system evolves in time under the Hamiltonian
(\ref{eq:H_T}) with $f(t)=f(0)$ for $t< 0$. At $t=0$ the force protocol is
started as usual. Although the formulation of FTs is independent of the value
of $f$ after $t=\tau$, it proves convenient to define $f(t)=f(\tau)$ for
$t>\tau$.

\section{Probability Distribution of Work and the FTs}

With these conceptual and technical preparations we now can define work
performed on the system in the time interval $[0, \tau]$ in the QBM model using
the trajectories given by the solutions of the Langevin equation
(\ref{eq:LanSol}) as:
\begin{equation}
\label{eq:def_work}
W=\int_0^{\tau} d t \frac{\partial H_T}{\partial t}=-\int_0^{\tau} dt
\dot{f}(t)X(t)=-\int_{-\infty}^{\infty} dt
\dot{f}(t)X(t)\equiv-\dot{f}^T\cdot X .
\end{equation}
In the last equality we utilized the notation of Eq.(\ref{def:dot}), where we
set the integration limits to plus and minus infinity. We will adopt this
convention for the rest of the paper. The superscript $T$ stands for transpose.
Since we have extended the range of integration in the redefined  $f(t)$ to the
entire real axis (see the end of the previous section) this change does not
introduce any error in the above equation.

We define the retarded Green's function as
$\gr(t-t')=g(t-t')\theta(t-t')$. Then for positive times:
\begin{eqnarray}
\label{eq:Xinf}
X(t)&=&[\gr \cdot f](t)+[\gr \cdot \xi](t), \\
\left \langle X(t)\right\rangle &=&[\gr \cdot f](t),\\
\sigma_{xx}(t,t') &\equiv&
\left \langle X(t)X(t')\right\rangle -\left
\langle X(t)\right\rangle \left \langle X(t')\right\rangle
=[\gr\cdot\hbar\nu\cdot\gr^T](t,t'),\\
\label{eq:Winf}
W&=&-\dot{f}^T\cdot\gr\cdot f-\dot{f}^T \cdot \gr\cdot\xi.
\end{eqnarray}
That $\sigma_{xx}(t,t')$ is a function of $t-t'$ only will be verified
explicitly later.

Work defined in Eq.(\ref{eq:Winf}) is linear in $\xi(t)$ and $\xi(t)$ is a
Gaussian random process. Thus  $W$ itself is a Gaussian random variable. As a
result the first two moments of $W$ specify its entire statistics given by:
\begin{eqnarray}
\label{eq:ForPro}
\cP(W)=\frac{1}{\sqrt{2 \pi \sigma_W^2}}e^{-(W-\left \langle
W\right\rangle )^2/2\sigma_W^2}.
\end{eqnarray}
The mean of work is given by:
\begin{eqnarray}
\label{eq:AveWor1}
\left \langle W\right\rangle &=&-\dot{f}^T\cdot \gr \cdot f.
\end{eqnarray}
Integrating this by parts and defining $\Delta
F=-(f(\tau)^2-f(0)^2)/2M\Omega^2$ we get:
\begin{eqnarray}
\label{eq:AveWor}
\left \langle W\right\rangle =\Delta F +\frac{\dot{f}^T \cdot h_e\cdot
\dot{f}}{2 M\Omega^2},
\end{eqnarray}
where we have defined $h_e(t,t')\equiv h(|t-t'|)$ and used the symmetry of the
integrand.
The standard deviation of work is calculated as:
\begin{eqnarray} 
\label{eq:std_work1}
\sigma_W^2 =\left \langle W^2\right\rangle -\left \langle W\right\rangle
^2=\dot{f}^T \cdot \sigma_{xx}\cdot \dot{f}.
\label{eq:StdAveRel}
\end{eqnarray}
Jarzynski equality states that:
\begin{eqnarray} 
\left \langle e^{-\beta W}\right\rangle =\int \rmd W \cP(W) e^{-\beta
W}=e^{-\beta \left(
\left \langle W\right\rangle -\beta \sigma_W ^2/2\right)}=e^{-\beta \Delta
F_T},
\end{eqnarray}
where $\Delta F_T$ is the difference in free energy of the combined
system for two different values of the external force $f$ calculated
quantum mechanically. Due to the linearity of the QBM model $\Delta
F_T$ has the same form as $\Delta F$ defined earlier, which is
the classical result. Note that this is only true for the difference of the
free
energies, since the quantum and classical free
energies themselves
are different even for the simple harmonic oscillator.
The quantum mechanical free energy in the case of $f=0$ is given by:
\begin{equation}
 F_T=F_B-\frac{1}{\beta} \log\left( \frac{1}{\beta \hbar \Omega}
\prod_{r=-\infty}^{\infty} \frac{\nu_r^2}{\Omega^2+\nu_r^2+2\nu_r
\hat{\gamma}(\nu_r)} \right)
\end{equation}
where $F_B$ is the free energy of the isolated bath. The corresponding free
energy in the classical model is simply the sum of the free energies of the
isolated system and bath.

 The equality of the difference of free energies due to a driving force
can be understood easily by noting that the main effect of the linear
driving force is to shift the energy levels. As a consequence Jarzynski
equality is satisfied if and only if:
\be
\label{eq:cond0}
\dot{f}\cdot\sigma_{\text{xx}}\cdot \dot{f} = \frac{\dot{f}\cdot h_e
\cdot \dot{f}}{\beta M \Omega^2}.
\ee
Note that this equality should hold for any $\dot{f}(t)$. 
This condition can be stated mathematically as 
\begin{eqnarray}
\frac{\delta}{\delta \dot{f}(s)} \int_{0}^{t_f} \rmd t \int_{0}^{t_f}
\rmd t' \dot{f}(t)\left[ \frac{h(|t-t'|)}{\beta M\Omega^2} -
\sigma_{xx}(t-t') \right]\dot{f}(t')=0,\\
 \int_{0}^{t_f} \rmd t  \dot{f}(t)\left[ \frac{h(|t-s|)}{\beta
 M\Omega^2} - \sigma_{xx}(t-s) \right]=0.
\end{eqnarray}
This equation should also be  valid for any $\dot{f}(t)$. Differentiating one
more time with respect to $\dot{f}(s')$ we get the condition:
\begin{eqnarray} 
\sigma_{xx}(s-s') = \frac{h(|s-s'|)}{\beta M\Omega^2}.
\label{eq:cond1}
\end{eqnarray}

For Crooks's fluctuation theorem we need to consider the reverse process which
corresponds to a reversed force protocol and an initial state with the force
value $f(\tau)$.  We will use a subscript R for the quantities associated with
the reversed process and no subscript for forward process.
\be
f_R(t)=f(\tau-t); \quad \Delta F_R=-\Delta F.
\ee
The corresponding work distribution is again specified by its first
two moments, which can be shown to be:
\begin{eqnarray}
\left \langle W\right\rangle _R&=&\left \langle W\right\rangle -2 \Delta F,\\
\nonumber
(\sigma_{W}^2)_R &=& \sigma_W^2.
\end{eqnarray}
Note that the standard deviation of work is the same for the forward
and reverse protocols. The probability distribution of work in the
reversed process is given by:
\begin{eqnarray} 
\label{eq:RevPro}
\cP_R(W)=\frac{1}{\sqrt{2 \pi
(\sigma_{W}^{2})_R}}e^{-(W-\left
\langle W\right\rangle _R)^2/2(\sigma_W^2)_R}=\frac{1}{\sqrt{2
\pi \sigma_{W}^{2}}}e^{-(W-(\left \langle W\right\rangle -2\Delta
F))^2/2\sigma_W^2}.
\end{eqnarray}
Consider the ratio:
\begin{equation} 
\frac{\cP_F(W)}{\cP_R(-W)}=e^{\frac{(\left \langle W\right\rangle -\Delta
F)}{\beta\sigma_W^2/2}\beta(W-\Delta F)}.
\end{equation}
Crook's fluctuation theorem is satisfied if
\begin{eqnarray} 
\label{eq:cond2}
\left \langle W\right\rangle -\Delta F=\beta\sigma_W^2/2.
\end{eqnarray}
As expected this condition is equivalent to the condition (\ref{eq:cond1}) for
the validity of Jarzynski equality.

Let us now try to understand the nature and meaning of condition
(\ref{eq:cond1}). $h$ and $g$ are solutions to the homogenous Langevin
equation. As such they do depend on the damping kernel but not on the
noise kernel. $\sigma_{xx}$ on the other hand depends on both the
damping kernel via $g$ and on the noise kernel. For this equality to
hold there has to be a relation between the noise and dissipation
kernels. The same conclusion can be reached by studying Eq.(\ref{eq:cond2}).
The average of work is independent of the noise kernel, but depends on the
damping kernel. On the other hand the standard deviation of work does depend on
both kernels.

There is indeed such a relationship between the damping and noise kernels: the
fluctuation dissipation relation (FDR). It is most easily presented in terms of
the Fourier transforms of the corresponding kernels.
\begin{equation}
\label{eq:QFDR}
\hbar\tilde{\nu}(\omega)=M \hbar \omega \coth(\beta \hbar \omega
/2)\tilde{\gamma}(\omega).
\end{equation}
However the quantum mechanical FDR in general does not satisfy condition
(\ref{eq:cond1}), and thus the FTs do not need to hold.
To see this note that the damping
kernel is
independent of $\hbar$. As a result the homogenous solutions of the Langevin
equation, $h$ and $g$, do not depend on $\hbar$. On the other hand
$\sigma_{xx}$ in general is a function of arbitrarily large powers of $\hbar$
via the coth term in the noise kernel.  FTs are satisfied if $\hbar$ is set to
zero. Corrections to FTs is expected at $O(\hbar^2)$.

\subsection{High and Low Temperature Regimes}

As  described in the previous subsection, noise kernel is the only
place where quantum effects are manifest, as can be seen by the
appearance of $\hbar$. Assumptions made on the properties of the bath
renders the quantum features associated with the initial state of the
system oscillator forgotten completely. In FTs the noise kernel appears only in
the standard deviation of work $\sigma^2_W$. In this subsection we will
investigate this term in the high and low temperature regimes.

Using the Fourier transform one can show from Eq.(\ref{eq:std_work1}) that:
\begin{equation}
\label{eq:std_work2}
\sigma_W^2=(2 \pi)^2\int_{-\infty}^{\infty}\rmd \omega
\tilde{f}_d(\omega) \tilde{\sigma}_{xx}(\omega)  \tilde{f}_d(-\omega),
\end{equation}
where $\tilde{f}_d(\omega)$ denotes the Fourier transform of $\dot{f}(t)$.
Recall that in our convention $\dot{f}(t)$ vanishes outside the interval
$[0,\tau]$, thus the Fourier transform is well-defined. Using the FDR
(\ref{eq:QFDR}) it can be shown that:
\be
\label{eq:sigma(omega)}
\tilde{\sigma }_{\text{xx}}(\omega )= \hbar  \omega  \coth
\left(\frac{\beta  \hbar  \omega }{2}\right)\frac{\tilde{h}_e(\omega
)}{2M \Omega ^2}.
\ee

\subsubsection{High temperature expansion}
\label{sec:hightemp}

For frequencies satisfying $\beta \hbar \omega< 1$, $\coth$ can be expanded
into a Laurent series:
\bea
\label{eq:laurent}
\coth(\frac{\beta \hbar \omega}{2})&=& \frac{2}{\beta \hbar
\omega}+\sum_{k=1}^{\infty}\frac{2^{2n} B_{2n}}{(2n)!}\left(
\frac{\beta \hbar \omega}{2} \right)^{2n+1},\\
\label{eq:hightempsigma(omega)}
\tilde{\sigma }_{\text{xx}}(\omega)&=&\frac{\tilde{h}_e(\omega
)}{\beta  M \Omega ^2}+\sum_{k=1}^{\infty}\frac{2^{2n}
B_{2n}}{(2n)!}\left( \frac{\beta \hbar \omega}{2} \right)^{2n+2}
\frac{\tilde{h}_e(\omega )}{ \beta M \Omega ^2},
\eea
where $B_n$ is the n'th Bernoulli number. If we assume that either
$\tilde{h}_e(\omega)$ or $\tilde{f}_d(\omega)$ decreases sufficiently
fast for large frequencies such that $\beta \hbar \omega\geq1$, the
Laurent series is a good expansion. Hence the characterization of
`high' temperature depends on two time scales: the intrinsic time
scale of the oscillator (determined by its interaction with the bath as
well as its natural frequency) and the time scale of the driving force. It is
reasonable to assume that $\tilde{h}_e(\omega)$ vanishes for
frequencies larger than the bath cutoff. Usually this is taken to be
very large. We will assume that $\tilde{f}_d(\omega)$ becomes
negligible at frequencies much smaller than this cutoff frequency, denoted as 
$\omega_h$. 
This is expected to be a reasonable assumption for typical driving forces. 
High temperature  refers to the condition $\beta \hbar \omega_h\ll 1$.

If we keep only the first term in the expansion (\ref{eq:hightempsigma(omega)})
we see that condition (\ref{eq:cond0}) for the validity of FTs is satisfied.
Deviations from FTs to all orders of $\hbar$ can be calculated to be:
\be
\frac{1}{ \beta M \Omega ^2}\sum_{n=1}^{\infty}\frac{2^{2n}
B_{2n}}{(2n)!}\left( \frac{\beta \hbar \imath}{2} \right)^{2n+2}
\dot{f}\cdot  h^{(2n+2)}_e \cdot \dot{f}.
\ee
The superscript on $h_e$ denotes the order of derivatives taken with respect to
the argument.  The correction term can also be written as:
\be
\label{eq:hightemp1}
\frac{1}{ \beta M \Omega ^2}\sum_{n=1}^{\infty}\frac{2^{2n}
B_{2n}}{(2n)!}\left( \frac{\beta \hbar}{2} \right)^{2n+2}
f^{(n+2)}\cdot  h_e \cdot f^{(n+2)}.
\ee
Note that the knowledge of the homogenous solution to the Langevin equation is
enough to calculate the correction term to all orders of $\hbar$.

\subsubsection{Low temperature expansion}
\label{sec:lowtemp}

Below we present the form of the standard deviation of work in a low
temperature expansion but we won't go into the details of the low temperature
expansion because the notion of trajectories will ultimately break down at
sufficiently low temperatures.
For high frequencies the following expansion of coth is more suitable
than Eq.(\ref{eq:laurent}):
\be
\coth\left(\frac{\beta \hbar
\omega}{2}\right)=\text{sgn}(\omega)\left[ 1+2\sum_{k=1}^\infty e^{- k
\beta \hbar |\omega|} \right],
\ee
\be
\label{eq:lowtemp1}
\tilde{\sigma }_{\text{xx}}(\omega )= \frac{\hbar}{2M \Omega ^2}
|\omega|  \tilde{h}_e(\omega )\left[ 1+2\sum_{k=1}^\infty e^{- k \beta
\hbar |\omega|} \right].
\ee
This expansion is convergent for all frequencies. However convergence
is fastest for $\beta \hbar \omega \gg 1$. If we assume that either
$\tilde{h}_e(\omega)$ or $\tilde{f}_d(\omega)$ decreases sufficiently
fast for $\omega \rightarrow 0$ such that $\beta \hbar \omega\leq1$,
expansion (\ref{eq:lowtemp1}) is a good one to use for Eq.(\ref{eq:std_work2}).
Hence the characterization of low temperature depends on two time scales: the
intrinsic time scale of the oscillator and the time scale of the driving force.
It is reasonable to assume that $\tilde{h}_e(\omega)$ vanishes for frequencies
lower than the lowest bath frequency. Usually this is taken to be very small.
We will assume that $\tilde{f}_d(\omega)$ becomes negligible at frequencies
much higher than the lowest bath frequency (This condition can be violated by a
very slowly changing driving force). Let us denote this frequency by
$\omega_l$. Low temperatures are defined by $\beta \hbar \omega_l\gg 1$.

\subsubsection{High temperature conditions and Markovian Dynamics}

An important special case is the Ohmic bath characterized by the spectral
density:
\begin{eqnarray}
J(\omega)=\frac{2M\g_0}{\pi}\omega.
\end{eqnarray}
Without a high frequency cutoff, the damping kernel becomes local in time. The
Langevin equation takes on the form:
\begin{eqnarray} 
\label{eq:MarLan}
M\ddot{X}(t)+2M\g_o\dot{X}(t)+M\Omega^2X(t)=f(t)+\xi(t).
\end{eqnarray}
Physically one would like to have a high frequency cutoff, which in turn makes
the damping kernel nonlocal in time. The high frequency cutoff also cures the
pathologies of the noise kernel that occur in the Ohmic case without cutoff. A
large cutoff $\Lambda$ ensures that the damping kernel is strongly peaked
around zero. If the driving force $f(t)$ doesn't change significantly on time
scales of order $1/\Lambda$, the Markovian approximation can be justified.

However, Markovian dynamics is not the criterion for FTs to be satisfied, high
temperature is. This is because even at high temperature if the bath is
non-Ohmic the dynamics of the open system can be non-Markovian.

\section{Relation between Classical and Quantum FTs} 

It is a well known fact that the dynamics of a quantum system with a quadratic
Hamiltonian is identical to the classical dynamics of the same model
\cite{Habib2004}. This applies to QBM model as well. One may wonder if FTs are
satisfied in classical dynamics, with the above observation, what is it then
that causes the possible violation of QFTs at low temperatures? Although the
dynamics is the same for quantum and classical models, initial conditions are
not. The thermal state at low temperatures is different for both. The damping
kernel does not depend on  the initial conditions and thus is the same for both
quantum and classical models. The noise kernel on the other hand depends on the
initial state of the bath. As a result it is the noise kernel that is different
and could give rise to deviations from FTs.

In the previous section we have seen how the classical limit is reached at high
temperatures. We identified high temperatures as the ones such that all the
relevant bath modes are multiply occupied. By relevant bath modes we mean those
that are within the range of frequencies of the external driving force. As is
well known from elementary quantum mechanics multiply-occupied harmonic
oscillators act classically. In this classical limit FTs are satisfied. 

Alternatively one can solve the classical version of the QBM model
exactly, which is possible due to the linearity of the model. Moreover
in the classical model one can use the thermal state of the combined
system instead of resorting to the infinite time
preparation\footnote{Like the quantum model, in the classical model
too the equivalence of both preparations can be proven exactly for any
spectral density.}. The result is a Langevin equation in which the
noise is correlated with the initial conditions of the system
oscillator. One can define a new noise which is uncorrelated to the
initial conditions of the system oscillator. This redefinition also
gets rid of the slip term \cite{FlemingAnnPhys, Weiss2008} in the Langevin
equation and one obtains the familiar form:
\begin{eqnarray} 
\label{eq:Lan}
M\ddot{X}(t)+2M\int_{0}^{t}\rmd s
\g(t-s)\dot{X}(s)+M\Omega^2X(t) &= f(t)+\xi(t),\\
\label{eq:ximean}
\lbr \xi(t)\rbr  &= 0,  \phantom{\g \g \g \g \g\g\g\g}\\
\label{eq:flucdiss}
\lbr \xi(t)\xi(t')\rbr &= \frac{2M}{\beta}\g(t-t'),\\
\label{eq:xicorr}
\lbr X_i \xi(t) \rbr &= \lbr P_i \xi(t) \rbr=0,
\end{eqnarray}
where the initial conditions are sampled from the reduced phase space density
of the system that is obtained from the thermal phase space density of the
combined system by integrating out the bath degrees of freedom.

Eqs.(\ref{eq:Lan}-\ref{eq:xicorr}) are the beginning point of the analysis of
\cite{Mai2007}. The authors of that paper start with the phenomenological
Langevin equation that is identical to (\ref{eq:Lan}). They further assume a
Gaussian noise with the classical FDR (\ref{eq:flucdiss}). Finally they assume
that the initial values of the system oscillator coordinates are sampled from
the classical phase space density $f_S(X_i,P_i,t_i)\propto\exp[-\beta
H_S(X_i,P_i,t_i)]$. This last point can be justified from the microphysics
model:
\begin{eqnarray} 
\nonumber
f_S(X_i,P_i,t_i)&=&\prod_{n=1}^N \int \rmd x_{ni} \int \rmd p_{ni}
f(X_i,P_i;\{x_{ni}\},\{p_{ni}\};t_i)\\ \nonumber
&\propto &\exp[-\beta H_S]\prod_{n=1}^N \int \rmd x_{ni} \int \rmd
p_{ni} \exp\left\{ -\beta\left( \sum_{n=1}^N \left[ \frac{p_{ni}^2}{2
m_n}+\frac{1}{2} m_n \omega_n^2\left( x_{ni}-\frac{c_n}{m_n \omega_n^2}X_i
\right)^2 \right]
\right)  \right\}\\ \nonumber
&\propto&\exp[-\beta H_S(X_i,P_i,t_i)].
\end{eqnarray}
Similarly the change in free energy that appears in Jarzynski equality and
Crooks's fluctuation theorem is that of the combined system. However, the
construction of the coupling and the renormalization term makes it coincides
with that of the isolated system oscillator. This clever scheme
notwithstanding,  we point out that in their phenomenological approach 
\cite{Mai2007} the free energy difference is mistakenly interpreted as that of
the free oscillator, since there is not enough information to track down its
origin. This kind of ambiguity and disconnectedness often  found in the
phenomenological models in the literature heightens the importance and
advantage of using a first-principles approach based on micro-physics models,
as is adopted here.

Starting from a microscopic model we were able to recover all the features of
the phenomenological Langevin equation. From there on, using the same analysis
as in \cite{Mai2007} leads to the verification of FTs.  However, it is crucial
to make the following distinction: In the phenomenological theory there is no a
priori reason why FTs should hold because the open system dynamics is not
Hamiltonian. As a result one needs to show the validity of FTs explicitly. 
In our formalism, on the other hand, we start with a closed (system + bath)
Hamiltonian system in a thermal state (of the combined system). Hence all the
premises of the FTs are satisfied and one expects that they should hold. What
needs 
to be done is to verify them from explicit calculations. 

One might object to this claim by noting that an uncountably infinite bath is
required for the preparation described in section (\ref{sec:prep}). The proof
of FTs for close Hamiltonian systems utilizes the Liouville
theorem, for which we have seen only proofs for finite number of
degrees of freedom. 
In this sense our model, with infinite preparation time also doesn't trivially
satisfy FTs, and needs the explicit verification. On the other hand for finite
baths one can use the thermal state of the combined system at t=0 and then the
FTs follow trivially. This second procedure is very easy for the classical
model though  somewhat complicated yet still straightforward for the quantum
model. The important point is that the infinite time preparation is only
introduced for technical convenience. It can be argued that for any relevant
times $t> 0$ the effect of an infinite bath can be approximated arbitrarily
closely by a large but finite bath. Hence our results are insensitive to the
unphysical assumptions about the bath we made in our derivation.

It is worth mentioning that Speck and Seifert \cite{Speck2007} have shown that
the Jarzynski relation holds for general classical ergodic systems governed by
stochastic dynamics including non-Markovian processes. 
Ohkuma and Ohta \cite{Ohkuma2007} studied classical systems described by a
non-linear, non-Markovian Langevin equation with Gaussian colored noise. Both
of these works are more general than our work when applied to classical systems
because they are not restricted to linear models. On the other hand both adopt
a phenomenological approach without an underlying microscopic model, as we do.

\section{Discussion}

\subsection{Comparison to previous work}
As mentioned in section \ref{sec:howtodefinework} there seems to be a consensus
on how to define work in closed quantum systems \cite{Campisi2011a}. Work is
defined as the
difference of the energy of the closed system measured at two
different times. This method is less attractive when applied to open
systems (treating the system+environment as the closed system) since
it  involves measuring the energy of the combined system. Furthermore
work is restricted to the open system, and it is only a part of the  total
energy which involves also heat exchange with the bath. This can lead to big
errors if one calculates the work of the combined system since work is the
difference of two large numbers.

In this paper we the use the decoherent history conceptual framework to explain
how the notion of trajectories in a quantum system can be made viable and use
them to define work for open quantum systems. These quantities are likely to be
more easily accessible than the energy levels for practical purposes related to
experiments,  especially for
open quantum systems. The classical mechanical definition of work in terms of
trajectories is used in the formulation
of FTs.

Work operator is another route taken \cite{Tasaki2000} but there is no
satisfactory definition of work as an operator \cite{Talkner2007}. Besides, the
work operator approach does not place any limit on the range of validity of its
predictions. Using the environment-induced decoherence scheme we can assess how
strong the noise in the environment need be to provide sufficient decoherence
to warrant the use of trajectories so as to be able to define work in open
quantum systems.  The parameter range whereby this condition is not satisfied
is likely related to the range where FTs may not hold quantum mechanically. 
The question of whether deviations from FTs can be observed in low temperature
experiments at all, and if so in which parameter range, requires more
quantitative analysis. This will be treated in a sequel paper.

Comparing with previous work in the literature the approach of
\cite{Chernyak2004} is closest to ours in spirit. However, in substance our
approach differs from theirs in several important ways,  as numerated below. 
Foremost a theoretical justification of the use of and the derivation of the
range of validity of
the trajectory concept in quantum mechanics are necessary in the formulation
of FTs.  To this end the authors of \cite{Chernyak2004} invoke continuous
measurements and wave function
collapse together with taking the semi-classical limit. We point out the key
conceptual and procedural steps which we believe \cite{Chernyak2004} are
flawed.

\paragraph{Conceptual flaws}

It is said  in \cite{Chernyak2004} that ``the classical limit can be reproduced
by using the Wigner function``. 
Also, ``$Q_+$ (Our $X$) is a classical coordinate variable and $Q_-$ (our $y$)
is a quantum coordinate"

These wrong statements stem from, we believe, a lack of understanding
of the central issues in quantum decoherence. Misconceptions like
these were common but were addressed and clarified in the 90s. See
e.g., \cite{Habib1990,Gell-Mann1993}.

\paragraph{Quantitative differences}

The range of validity is not stated clearly in \cite{Chernyak2004}
and the generating functional of work given in their Eq.(9) is said to be valid
at arbitrary temperature. We believe this is an overclaim.

In the {\it dechis} or {\it envdec} formalism trajectories emerge due to the
influence of the environment, in particular, the strength of noise: The
stronger the noise the more pronounced trajectories take shape, the weaker the
noise, the more quantum features prevail.  These conditions of classicality can
be quantified clearly and from them one obtains the criteria for determining
the range of validity of quantum FTs as we discussed in an earlier section.

Eq.(13) of \cite{Chernyak2004} gives the lowest order in $\hbar$
correction to the Jarzynski equality. We provide the corrections to
arbitrary orders of $\hbar$ in our Eq.(\ref{eq:hightemp1}) in terms of the
homogenous solutions
to the Langevin equation. Furthermore we show that these corrections apply to
both Crooks's fluctuation
theorem as well as Jarzynski's equality. At the classical level we derive
Crooks's
fluctuation theorem and Jarzynski's equality for the Brownian motion
model. 

\subsection{New issues brought forth}
\label{sec:newiss}

The {\it dechis} and {\it envdec} approach bring forth a number of new
issues which were not so clearly noted before. We name three here.

\paragraph{Initial state preparation }

Initial state preparation is an important aspect of FTs. Most of the literature
on
FTs for closed systems is usually clear on this aspect. However a certain level
of ambiguity exists in open system treatments. In this work we considered  an
initial thermal state for the closed system made up of the subsystem plus
its environment. However for computational ease and clarity of
exposition we developed an equivalent initial state preparation method based on
product initial states for the system and the bath.
Our initial state preparation replaces the system's dependence on the
initial state by  the properties of noise statistics. As a result our
preparation method has
only one probabilistic element as opposed to two. This makes the analysis
clearer and the identification of quantum effects easier.

\paragraph{On the meaning of the average in Eq.(\ref{eq:Jarz}).}

The averages that are calculated using the statistics of noise can
alternatively
be expressed in terms of expectation values of quantum mechanical operators.
The important point is that products of position and momentum operators need to
be symmetrized owing to the properties of Wigner function, which is used in the
averaging process. In the specific case of Jarzynski equality, we observe that
the average
over noise realizations
can also be obtained by taking the
expectation value of the quantum mechanical operators as:
\bea
\left \langle e^{-\beta W}\right\rangle &=&\int D\xi P[\xi] \int \rmd x_0 \rmd
p_0 W_r(x_0,p_0,0)
e^{\beta \int_0^\tau \rmd t \dot{f}(t) x_\xi(t)}\nonumber\\
&=& \tr_{S+B}\left[ e^{\beta \int_0^\tau \rmd t
\dot{f}(t)\hat{x}_H(t)}\hat{\rho}_\beta
\right].
\eea
In this special case symmetrization is achieved by the exponential
function together with the fact that the dynamics is linear and work
itself is a linear function of position. Consequently we don't need to
impose the symmetrization procedure explicitly. It is in this strict sense that
the results that are obtained using the work operator $\hat{W}\equiv
-\int_0^\tau \rmd t \dot{f}(t)
\hat{x}_H(t)=\hat{H}_H(\tau)-\hat{H}_H(0)$ for Jarzynski equality
agree with our results obtained via trajectories.

\paragraph{How to decide if possible violations to FTs can be observed?}

The formulation of FTs involves  averages over noise realizations,
with idealized situations where trajectories are perfectly well
resolved for each realization of noise.  But of course in an experiment, even
classically, there is only finite resolution. Let us assume that the resolution
of the experiment is independent of temperature. This introduces an error to
the FTs obtained from this data that is independent of the temperature.

In the quantum case the condition  $\sigma^2\sim \nu^{-1}$ suggests that for
stronger noise we can resolve the
trajectory to a higher precision. As the noise weakens such as at decreasing
temperature  the stochastic features of classical trajectory are enhanced and
measurement results on a particle's trajectory becomes less precise. Further
weakening the noise we will get to a point in which quantum or ``Heisenberg"
noise dominates \cite{Hu1995}. Here lies a fundamental difference between
classical and quantum. In quantum mechanics the ability of resolving
trajectories is not only determined by the precision of the measurement device
but also by the temperature. As a result the error in FTs introduced by the
resolution of trajectories increases constantly as the temperature is lowered,
unlike in classical mechanics. Below a certain temperature, upon entering the
quantum dominated regime, the imprecision in  measurements will become too
large to render any free energy calculations using FTs meaningless.

The properties of noise acting on the quantum Brownian particle are
different from the noise in the corresponding classical model, as was
shown above. This introduces a deviation from FTs which is independent
of the error introduced by the limited precision of measurements
(discussed in the previous paragraph). The quantum corrections to the
noise kernel become larger at lower temperatures. As a result we
expect to observe deviations from FTs at low temperatures. 
However, as we learned before, trajectories are not well-defined at
arbitrarily low temperatures. Further quantitative analysis is necessary to
establish the domain of validity of our
approach and the magnitude of possible violations to FTs within this domain as
a function of temperature.

\subsection{Work in progress} We mention two aspects which command our current
attention on this problem.

a.  We want to be able to answer the following question: Is
there a temperature range for which experiments done to the precision
prescribed by the decoherent histories interpretation give answers
different from the corresponding calculations based on classical
mechanics with the same measurement precision imposed (theoretically
measurements in  classical systems can have infinite precision)? If so
what are the form/size of these violations? Such a difference would be due to
quantum mechanics exclusively and this is what we mean by violations of the FTs
at low temperatures due to quantum
mechanical effects. 

b. Unlike in other approaches to quantum FTs, the use of open quantum system 
concepts and especially the influence functional method adopted here enable us
to define and quantify heat flow in terms of the dissipative dynamics of the
open system which results from a self-consistent treatment of the back-action
from its environment. We want to take advantage of this approach to address
questions about energy exchange between the system and the bath, where
applications abound. \\

\noindent{\bf Acknowledgments} 
We thank Prof. C. Jarzynski for informative discussions and helpful comments on
a preliminary draft. 
YS acknowledges useful discussions with Chris Fleming on the initial state
preparation and with Ryan Behunin on the worldline path integral formalism. 

\bibliography{FTs}

\end{document}